\documentclass[aps,prl,twocolumn,nofootinbib,raggedbottom]{revtex4-2}

\usepackage[usenames,dvipsnames]{xcolor}
\usepackage{hyperref}
\hypersetup{
	colorlinks=true,
	urlcolor=Maroon,
	linkcolor=RoyalBlue,
	citecolor=Maroon,
	bookmarks=true,
	pdftitle={Scattering with Neural Operators},
	pdfauthor={Sebastian Mizera},
	pdfdisplaydoctitle=true,
	pdfstartview=FitH
}

\usepackage{amsmath}
\usepackage{amsfonts}
\usepackage{amssymb}
\usepackage{mathtools}
\usepackage{microtype}
\usepackage{bm}
\usepackage[utf8]{inputenc}
\usepackage{blkarray}
\usepackage{bigstrut}
\usepackage{nccmath}
\usepackage{enumitem}
\usepackage{braket}
\usepackage{dsfont}
\usepackage[export]{adjustbox}
\usepackage[english]{babel}
\usepackage{stmaryrd}
\usepackage{orcidlink}

\usepackage{caption}
\makeatletter
\newcommand{\xMapsto}[2][]{\ext@arrow 0599{\Mapstofill@}{#1}{#2}}
\def\Mapstofill@{\arrowfill@{\Mapstochar\Relbar}\Relbar\Rightarrow}
\makeatother

\allowdisplaybreaks
\graphicspath{{figures/}}

\renewcommand{\Re}{\mathrm{Re}}
\renewcommand{\Im}{\mathrm{Im}}

\def \be {\begin{equation}}
\def \ee {\end{equation}}

\def \C  {\mathbb{C}}

\def \R  {\mathbb{R}}

\def \e  {\mathrm{e}}
\def \d  {\mathrm{d}}
\def \N  {\mathcal{N}}
\def \F  {\mathcal{F}}
\def \I  {\mathbf{I}}
\def \H  {\mathbf{H}}
\def \O  {\mathbf{O}}
\def \L  {\mathcal{L}}
\def \P  {\mathcal{P}}
\def \K  {\mathcal{K}}
\def \GT {\mathrm{GT}}
\def \NO {\mathrm{NO}}

\definecolor{yellowish}{rgb}{0.880722,0.611041,0.142051}
 
\begin{document}

\title{Scattering with Neural Operators}

\author{Sebastian Mizera \!\orcidlink{0000-0002-8066-5891}}
\affiliation{Institute for Advanced Study, Einstein Drive, Princeton, NJ 08540, USA}

\begin{abstract}\noindent
Recent advances in machine learning establish the ability of certain neural-network architectures called \emph{neural operators} to approximate maps between function spaces.
Motivated by a prospect of employing them in fundamental physics, we examine applications to scattering processes in quantum mechanics. We use an iterated variant of Fourier neural operators to learn the physics of Schr\"odinger operators, which map from the space of initial wave functions and potentials to the final wave functions.
These deep operator learning ideas are put to test in two concrete out-of-distribution problems: a neural operator predicting the time evolution of a wave packet scattering off a central potential in $1+1$ dimensions, and the double-slit experiment in $2+1$ dimensions. At inference, neural operators can become orders of magnitude more efficient compared to traditional finite-difference solvers.
\end{abstract}

\maketitle

\section{Do Androids Dream\\ of Schrödinger's Cats?}

Recent advances in machine learning demonstrated the ability of neural networks to approximate not only functions, but also non-linear operators, using various architectures collectively known as \emph{neural operators} \cite{Lu_2021,anandkumar2019neural,li2020fourier,kovachki2021neural,LU2022114778}. In this work, we ask whether they can serve as a practical computational tool in fundamental physics.\footnote{Neural networks, but not operators, have already had transformative impact on particle-physics phenomenology and beyond, see, e.g., \cite{Carleo:2019ptp,Karagiorgi:2021ngt,Butter:2022rso}.} This question is motivated in part by the exploding complexity of perturbative quantum field theory computations needed for precision predictions at the Large Hadron Collider \cite{Caola:2022ayt,FebresCordero:2022psq}, calling for rethinking how to represent and compute S-matrix elements efficiently.

This investigation has to start somewhere and here we consider arguably the simplest scattering operator in $(d+1)$-dimensional quantum mechanics,
\be
S[V(\vec{x})] = \mathcal{T} \e^{-i/\hbar \int_{0}^{T} \hat{H}[V(\vec{x})]\, \d t}\, ,	
\ee
which acts on the initial position-space wave function $\Psi(\vec{x},0)$ to produce the final one $\Psi(\vec{x},T)$ at some fixed time $T$:
\be\label{eq:S}
\Psi(\vec{x},T) = S[V(\vec{x})]\, \Psi(\vec{x},0)\, .
\ee
Here, $(\vec{x},t)$ are the space-time coordinates,  ${\cal T}$ is the time-ordering operator, and $\hat{H} = -\tfrac{\hbar^2}{2m} \nabla^2 + g V(\vec{x})$ is the time-independent Hamiltonian, where $m$ is the particle mass, and $g$ is a coupling constant. We emphasized the functional dependence on the potential $V$, which is precisely what makes $S$ a non-linear operator, viewed as acting simultaneously on the space of $V$'s and initial conditions for $\Psi$.

We ask whether, instead of computing the time evolution using traditional methods, the Schr\"odinger operator $S$ can be represented as a neural operator ${\cal N}$ (defined more precisely below), so that:
\be\label{eq:N}
\Psi(\vec{x},T) \stackrel{?}{=} {\cal N}[V(\vec{x}), \Psi(\vec{x},0)]\, .
\ee
The \emph{universal approximation theorem} for operators, originally due to Chen and Chen \cite{392253}, actually guarantees an affirmative answer. Translated to our setup, it implies that for any $\epsilon > 0$, there exists a complicated enough ${\cal N}_{\epsilon}$ such that the $L^2$-norm $\| \Psi(\vec{x},T) - {\cal N}_{\epsilon}[V(\vec{x}), \Psi(\vec{x},0)] \| < \epsilon$ for every continuous $V(\vec{x})$ and $\Psi(\vec{x},0)$, see \cite{10.1093/imatrm/tnac001,10.5555/3546258.3546548,deryck2022generic} for details. We will make this statement more precise shortly.

Nevertheless, it is not known how to find such an ${\cal N}_{\epsilon}$ constructively (it would be akin to looking for the next best-selling novel in the decimal expansion of $\pi$).
Indeed, constructing decent $\N$'s became practically viable only recently, by combining neural operators with deep learning ideas \cite{Lu_2021,kovachki2021neural}. In this work, we examine whether this framework allows us to machine learn the Schr\"odinger operator in a meaningful way.

\begin{figure}
\centering
\includegraphics[width=\columnwidth]{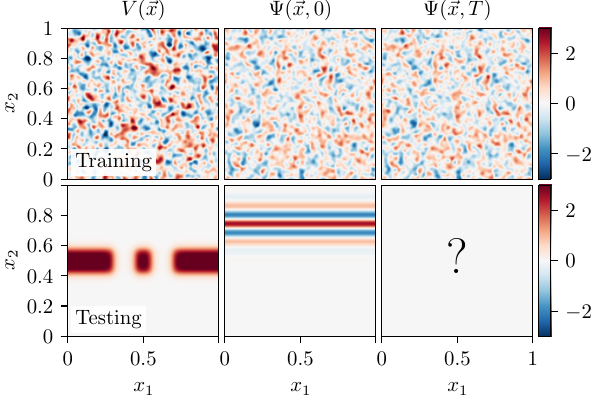}
\caption{\label{fig:intro}Example training (top) and testing (bottom) data with the input in the first two columns and the output in the third one. Only real parts are displayed for conciseness.}
\end{figure}

\paragraph{\normalfont\textbf{Strategy}}
Let us sketch the general idea first. The neural operator is trained on triples of data:
\be
\underbrace{V(\vec{x}),\; \Psi(\vec{x},0)}_{\mathrm{input}},\; \underbrace{\Psi(\vec{x}, T)}_{\mathrm{output}} \, ,
\ee
where the input is a Gaussian noise and $\Psi(\vec{x}, T)$ is computed exactly according to \eqref{eq:S}, see Fig.~\ref{fig:intro} (top). The learning process is entirely data-driven and has no physics input: ${\cal N}$ has to effectively ``dream up'' its own representation of the Schr\"odinger equation from scratch. In fact, the neural operator learns not just one instance of this equation, but instead a whole family parameterized by the interaction potential $V$.

It would not be surprising if $\N$ could accurately solve problems drawn from the same distribution as it was trained on. Therefore, after training, or \emph{at inference}, $\N$ is instead tasked with predicting the time evolution of a previously-unseen problem: for example, a wave packet approaching a double-slit potential, see Fig.~\ref{fig:intro} (bottom). In other words, we are probing the \emph{generalization} or \emph{out-of-distribution error}: the ability of neural operators to extrapolate the information learned during training to new problems.

The prediction needs to be accurate enough to be able to iterate this process $k$ times and study strongly-coupled scattering by evolving the wave function $\Psi(\vec{x}, k T)$ for long times.
As we will see in Tab.~\ref{tab:performance}, at inference, this way of encoding physics can become faster and more memory-efficient compared to traditional algorithms, such as finite-difference methods, because once the operator $\N$ is learned, it does not need to be recomputed for every new input $V$ and $\Psi$.

\section{Architecture}

We use a version of the Fourier neural operator (FNO) architecture \cite{li2020fourier} adapted to the problem at hand.

\paragraph{\normalfont\textbf{Channels}} The data is organized into \emph{channels}. We take the input $\I\!: D \to \R^{d+3}$ to consist of $d{+}3$ channels,
\be\label{eq:I}
\I(\vec{x}) = \Big( V(\vec{x}),\; \Re\, \Psi(\vec{x},0),\; \Im\, \Psi(\vec{x},0),\; \vec{x} \Big) \, ,
\ee
where $D \ni \vec{x}$ is the spatial domain we take to be the unit torus $D = \mathbb{T}^d$ (we impose periodic boundary conditions for simplicity). In practice, one samples points from a discretization of $D$, such as a uniform lattice of $N^d$ points. Likewise, the output $\O\!: D \to \R^2$ consists of $2$ channels:
\be
\O(\vec{x}) = \Big( \Re\, \Psi(\vec{x}, T),\; \Im\, \Psi(\vec{x}, T) \Big)\, .
\ee
Splitting $\Psi$'s into real/imaginary parts and adding \emph{positional embeddings} $\vec{x}$ is used to increase numerical stability and facilitate the learning process.

The architecture will be designed to be independent of the specific discretization of $D$, which in particular means that a neural operator trained on a coarse lattice can be used to make predictions on a finer one, as will be discussed below.

\paragraph{\normalfont\textbf{Neural operator}} A neural operator $\N$ is written as a composition of $\mathrm{L}{+}2$ \emph{layers} resembling standard deep neural networks \cite{kovachki2021neural}:
\be
\N = \mathcal{P} \circ f_{\mathrm{L}} \circ \cdots \circ f_2 \circ f_1 \circ \mathcal{L}\, .
\ee
Here, $\mathcal{L}$ and $\mathcal{P}$ are the \emph{lifting} and \emph{projection} layers respectively, which act as identities in $D$ and whose role is to map the input data $\I$ to its hidden representations with $h$ channels, $\H_{\ell}\!: D \to \R^h$ for $\ell=0,1,\ldots,\mathrm{L}$, and then back to the output $\O$:
\be
\I \xmapsto{\L} \H_0 \xmapsto{f_1} \H_1 \xmapsto{f_2} \cdots \xmapsto{f_{\mathrm{L}}} \H_{\mathrm{L}} \xmapsto{\P} \O\, .
\ee
For us, $\L$ and $\P$ are implemented as $1$- and $2$-layer perceptrons respectively, with the hidden dimension $p$ in the second case. Typically, $\mathrm{L}=4$ and $h, p \gg d$. The hidden-layer functions $f_{\ell+1}$ are defined through
\be
\H_{\ell+1}(\vec{x}) = \sigma \Big( W_\ell \H_\ell (\vec{x}) + (\K_\ell \H_{\ell} )(\vec{x}) + b_\ell \Big) + s_\ell \H_{\ell}(\vec{x})  \, ,
\ee
where the \emph{weights} $W_\ell \in \R^{h \times h}$, \emph{biases} $b_\ell \in \R^h$, and \emph{skip connections} $s_\ell \in \R$ act as identities in $D$. The activation function $\sigma$ is the only non-linearity and is applied element-wise. We take it to be the Gaussian Error Linear Unit (GELU) \cite{hendrycks2023gaussian}. 

\begin{figure}[t]
	\includegraphics[scale=1]{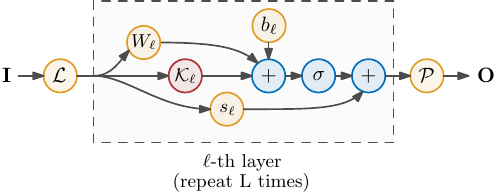}
	\caption{\label{fig:architecture}Diagram of the neural operator architecture used in this work. The Fourier layer (red) is introduced in \eqref{eq:Fourier-layer}.}
\end{figure}

\paragraph{\normalfont\textbf{Fourier layers}} The key objects are the (linear) integral kernel operators $\K_\ell$ with $\ell=1,2,\ldots,\mathrm{L}$, which take the general form
\be\label{eq:KH}
(\K_\ell \H_\ell)(\vec{x}) = \int_D\!\! K_\ell(\vec{x},\vec{y})\, \H_{\ell}(\vec{y})\, \d^d \vec{y}
\ee
for all $\vec{x} \in D$. Note this is the only layer acting non-diagonally in $D$.

Out of the previously-explored choices for parametrizing the convolution kernel $K_\ell$ \cite{anandkumar2019neural,li2020fourier,kovachki2021neural}, we write \eqref{eq:KH} using Fourier transforms $\F$ and their inverses $\F^{-1}$:
\be\label{eq:Fourier-layer}
\K_\ell \H_\ell = \F^{-1} \!\left[ \F[K_\ell]\, \F[\H_\ell] \right]\, .
\ee
This choice allows us to represent $\F[K_\ell]$ directly in Fourier space $D^\ast \ni \vec{p}$, as a tensor $\mathbf{T}_\ell \in \C^{F\times h\times h}$, where $F$ is the number of Fourier modes. The expression \eqref{eq:Fourier-layer} can then be evaluated efficiently using fast Fourier transforms and tensor multiplication.

To summarize, the parameters of $\N$ are the entries of $W_\ell$, $\mathbf{T}_\ell$, $b_\ell$, and $s_\ell$ that we will optimize for, see Fig.~\ref{fig:architecture}. The hyperparameters are all the numbers describing the network properties, such as $\mathrm{L}$, $h$, $F$, etc. In this case, the aforementioned universal approximation theorem states that one can always find large enough values of the hyperparameters such that there exist specific $W_\ell$, $\mathbf{T}_\ell$, $b_\ell$, and $s_\ell$ for which the $L^2$-error is bounded by $\epsilon$ for any input in a compact subset of a Banach space \cite{10.1093/imatrm/tnac001,10.5555/3546258.3546548,deryck2022generic}. Note that the theorem would not be true if we did not include the non-linearity $\sigma$ and the Fourier layers.

In practice, we truncate the number of modes to be half along each dimension, i.e., $F = (N/2)^d$, and ensure reality of the output by imposing $\F[K_\ell](\vec{p}) = \F[K_\ell](-\vec{p})^\ast$ on the remaining ones. 
Since the tensors $\mathbf{T}_\ell$ are by far the largest part of the neural operator (contributing $2\mathrm{L}m h^2$ real parameters), it pays off to instead represent them using Tucker decomposition with rank, or compression ratio, $r$ \cite{kossaifi2023multigrid}.

We have tried adding a number of other features, including normalization layers, dropout regularization \cite{10.5555/2627435.2670313}, and Sobolev training \cite{czarnecki2017sobolev} without drastic improvements in the results. Nonetheless, we suspect they will be important in large-scale applications of neural operators.

There exist different open-source implementations of FNO \cite{github_neural_operators}, as well as other neural-operator architectures, including DeepONet \cite{Lu_2021} and their variations \cite{li2020multipole,gupta2021multiwaveletbased,bhattacharya2021model,guibas2022adaptive,cao2021choose,rahman2023uno,fanaskov2022spectral,doi:10.1137/22M1477751,ZHU2023116300,jiang2023fouriermionet,doi:10.1126/sciadv.abi8605,PhysRevResearch.4.023210,CAI2021110296,ZHU2023116064,li2022fourier,deng2021convergence,ovadia2023ditto,rahman2022generative}.

\begin{figure}
	\centering
	\includegraphics[width=0.95\columnwidth]{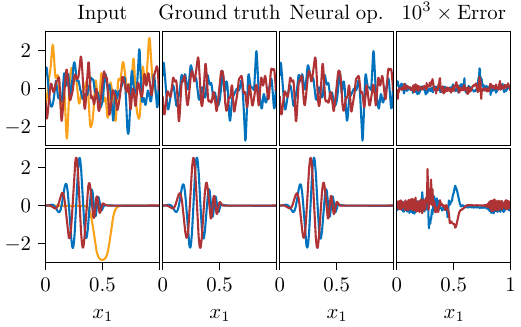}
	\caption{\label{fig:single-step}Comparison between the exact and inferred wave functions for training (top) and testing (bottom) datasets in $d=1$. \textbf{From left to right:} Potential $V(x_1)$ (\textcolor{yellowish!90!black}{yellow}) and the initial wave function $\Psi(x_1,0)$ (real part in \textcolor{Maroon}{red}, imaginary part in \textcolor{RoyalBlue}{blue}); exact wave function $\Psi_{\GT}(x_1, T)$; neural operator output $\Psi_{\NO}(x_1, T)$; and the error $\Psi_{\GT}(x_1, T) - \Psi_{\NO}(x_1, T)$ magnified $10^3 \times$ for visibility. The difference between $t=0$ and $t=T$ is non-zero, but barely visible by eye due to the choice of a small time step; see Fig.~\ref{fig:wave-packet} for time evolution across $t \leqslant 256 T$.}
\end{figure}

\paragraph{\normalfont\textbf{Loss function}} We use the $L^2$-error as the loss function:
\be
\mathrm{loss} = \Big\| \Psi(\vec{x},T) - \N[V(\vec{x}), \Psi(\vec{x},0)]\, \Big\|\, ,
\ee 
where $\Psi(\vec{x},T)$ is the answer computed using the Crank--Nicolson finite difference method, which for our purposes will be referred to as \emph{exact} or as the \emph{ground truth} below. To avoid confusion, we will add subscripts: $\Psi_{\GT}(\vec{x},T) = \Psi(\vec{x}, T)$ and $\Psi_{\NO}(\vec{x},T) = \N[V(\vec{x}), \Psi(\vec{x},0)]$ to distinguish between the exact and inferred computations.

We use the AdamW optimizer \cite{loshchilov2018decoupled}, which is an adaptive variant of stochastic gradient descent, with learning rate $\nu$ and weight decay $w$, together with a scheduler halving $\nu$ every time the training loss reaches a plateau. 

\begin{figure}
	\centering
	\includegraphics[width=0.8\columnwidth]{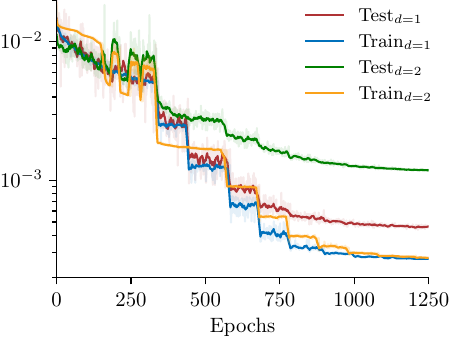}
	\caption{\label{fig:learning-curves} Example training and testing losses for $d=1,2$. Single epoch takes $\sim 1.8$ ($d=1$) and $\sim 4.9$ seconds ($d=2$).}
\end{figure}

\paragraph{\normalfont\textbf{Training dataset}}
The training set is constructed by drawing the first three entries of \eqref{eq:I} independently from a Gaussian random field with spatial width $\mu = 0.1$ (i.e., the power spectrum $\propto \e^{- \mu^2 |\vec{p}|^2/2}$), sampled over a uniform grid of $N^d$ points on $\mathbb{T}^d$. In particular, the potential $V(\vec{x})$ is purely real. In order to probe unitarity, each sample is normalized such that $\| \Psi(\vec{x},0) \| = 1$ and we similarly set $\| V(\vec{x}) \| = 1$. We then create a \emph{timeline} by iteratively solving for $\Psi_{\GT}(\vec{x}, kT)$ up to $k \leqslant M = 256$. Hence, each timeline produces $M{-}1$ input-output pairs. For numerical stability, we normalize each output such that $\| \Psi_{\GT}(\vec{x}, kT) \| = 1$ before evolving it to $\Psi_{\GT}(\vec{x}, (k{+}1)T)$. We repeat this process $K = 32$ times, which results in the training dataset of size $n_{\mathrm{train}} = K(M{-}1) = 8160$. See Fig.~\ref{fig:intro} and \ref{fig:single-step} (top rows) for examples.

We emphasize that it is possible to achieve much better performance with a training set adapted to scattering problems, for example by including $\Psi$'s resembling wave packets. Likewise, one can include probability conservation in the loss function, hard-code the fact $\N$ is supposed to be linear in $\Psi$, or make use of other physics-informed training strategies \cite{RAISSI2019686,goswami2022physicsinformed,li2023physicsinformed,Rotman:2022hzi}. Here, we purposely do not employ these steps, because our goal is to probe the out-of-distribution error and find out whether $\N$ has learned quantum physics.

All computations are performed on a single NVIDIA A100 GPU with 40GB memory.

\section{Wave Packet Scattering}

As the first application, we consider wave packet scattering in $d=1$. We set $\hbar=m=1$ and $g= 3{\times} 10^3$. In those units, the time step is chosen to be $T = 6.3 {\times} 10^{-5}$, which is fine enough to produce a ``movie'', see \cite{supplemental}. Much smaller $T$'s can cause numerical approximation errors and much larger ones prove more difficult to train. The spatial dimension is a unit circle $\mathbb{T}^1 \ni x_1$ and we discretize it with $N = 256$ points.

\paragraph{\normalfont\textbf{Testing dataset}}
The testing dataset is prepared in exactly the same way as the training one, except it is out-of-distribution: we start with the specific wave packet $\Psi(x_1,0)$ approaching a potential well $V(x_1)$, as illustrated in Fig.~\ref{fig:single-step} (bottom left). After evolving it through time $t \leqslant M T$, the testing dataset comprises of $n_{\mathrm{test}} = M{-}1 = 255$ samples.

\begin{figure}
	\centering
	\includegraphics[width=0.95\columnwidth]{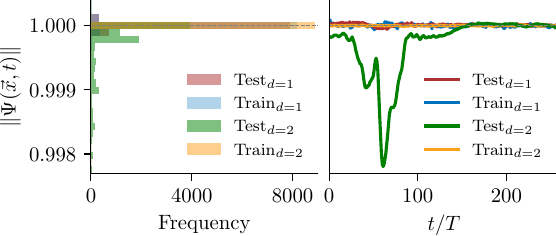}
	\caption{\label{fig:unitarity}Unitarity of the neural operator outputs for the training and testing datasets in $d=1,2$. \textbf{Left:} Histogram of $\| \Psi_{\NO}(\vec{x},t) \|$ after one step $t=T$. \textbf{Right:} Norms $\| \Psi_{\NO}(\vec{x},t) \|$ along the timeline $t \leqslant 256T$.}
\end{figure}

\paragraph{\normalfont\textbf{Hyperparameters}}
We performed a Bayesian search over the hyperparameter space, which revealed preference for relatively small networks that are less prone to overfitting. In the results below, we use $h=64$, $p=512$ hidden and projection channels, $\mathrm{L} = 4$ Fourier layers, and Tucker rank $r = 10^{-2}$. Learning rate is set to $\nu = 10^{-3}$, weight decay to $w = 10^{-5}$, and training is performed in batches of size $b = 32$. The resulting model has $\sim 9 {\times} 10^{4}$ parameters. 

Example learning curves are shown in Fig.~\ref{fig:learning-curves}, where we also keep track of the testing loss as a benchmark. At first, it closely follows the training loss, but around epoch $\gtrsim 500$ the neural operator starts overtraining (memorizing instead of learning). We still find it beneficial to continue learning until both losses stabilize with the training loss converging to $\sim 3 {\times} 10^{-4}$ and testing to $\sim 5 {\times} 10^{-4}$. The kinks in the losses occur in places where the learning rate $\nu$ gets halved.
Example training and testing data samples are compared in Fig.~\ref{fig:single-step}. Note that in the latter case, the wave function has flat near-zero segments, which are very non-generic from the perspective of the training data, but nevertheless correctly evolved.

\paragraph{\normalfont\textbf{Unitarity}}
We found that unitarity can be treated as a proxy for the confidence of the neural operator about its predictions. In Fig.~\ref{fig:unitarity} (left), we display a histogram of $\| \Psi_{\NO}(x_1, T) \|$ over the whole training and testing datasets. It demonstrates the neural operator has learned unitarity with $\sim 10^{-4}$ precision.

\begin{figure}
	\centering
	\includegraphics[width=\columnwidth]{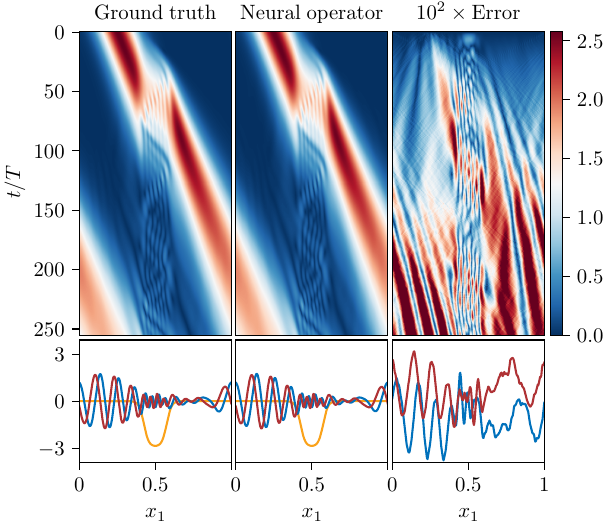}
	\caption{\label{fig:wave-packet} Wave packet scattering in $d=1$. \textbf{Top:} Absolute values of the exact computation $| \Psi_{\GT}(x_1,t)|$; the neural operator output $| \Psi_{\NO}(x_1,t)|$; and the error $| \Psi_{\GT}(x_1,t) - \Psi_{\NO}(x_1,t)|$ multiplied by $10^2$. \textbf{Bottom:} Final-step wave function at $t=256T$ in the same notation as in Fig.~\ref{fig:single-step}.\vspace{-1em}}
\end{figure}

\paragraph{\normalfont\textbf{Long-term predictions}}
Finally, we discuss the ability of the neural operator to make long-term scattering predictions by iterating $\N$,
\be
\Psi_{\NO}(\vec{x}, kT) = \N^{k}[V(\vec{x}), \Psi(\vec{x},0)]\, ,
\ee
where $k \leqslant M = 256$. For simplicity of exposition, we keep the input potential $V$ constant, even though the same $\N^k$ is also capable of treating time-dependent $V$'s. As with the finite-difference methods, we normalize $\| \Psi_{\NO}(\vec{x}, kT)\| = 1$ after each iteration. Example timeline of $\| \Psi_{\NO}(\vec{x}, t)\|$ before normalization are shown for $t \leqslant 256T$ in Fig.~\ref{fig:unitarity} (right) for examples of training and testing samples. They remain accurate to within $\sim 10^{-4}$ of identity, indicating high confidence in reliable predictions.

A full timeline is presented in Fig.~\ref{fig:wave-packet}, which shows the wave packet approaching the central potential, interfering with it, and emerging on the other side with a time delay and a spreading effect. The error builds up over time, but stays within $\sim 0.02$ even in the final time step $t = 256 T$, showing strong generalization capabilities. It demonstrates that the iterated neural operator $\N^k$ has effectively learned the physics of the Schr\"odinger operator at strong coupling in $d=1$.

\section{Double-Slit Experiment}

We next discuss neural operators in $d=2$, with the out-of-distribution test problem being a wave packet scattering off a double-slit potential, see Fig.~\ref{fig:intro} (bottom) for the initial conditions.

The hyperparameters used are the same as in the $d=1$ case, except for the Tucker rank, which we take to be $r=10^{-3}$ and the grid size, $N=64$. The resulting number of parameters is $\sim 1.2 {\times} 10^{5}$. Training and testing dataset sizes remain the same as in $d=1$.

\paragraph{\normalfont\textbf{Learning}}
As expected, the neural operator becomes more difficult to train in $d=2$, as exemplified by the learning and unitarity curves in Fig.~\ref{fig:learning-curves} and \ref{fig:unitarity} respectively. The generalization error is worse by around an order of magnitude compared to $d=1$: the unitarity errors stay within $\sim 4{\times} 10^{-3}$ of identity; the training and testing losses reach $\sim 3 {\times} 10^{-4}$ and $\sim 1.1 {\times} 10^{-3}$ respectively. Unitarity dips around the time when the wave function develops high-frequency modes due to encountering the potential barrier.

\begin{figure}[t]
	\centering
	\includegraphics[width=\columnwidth]{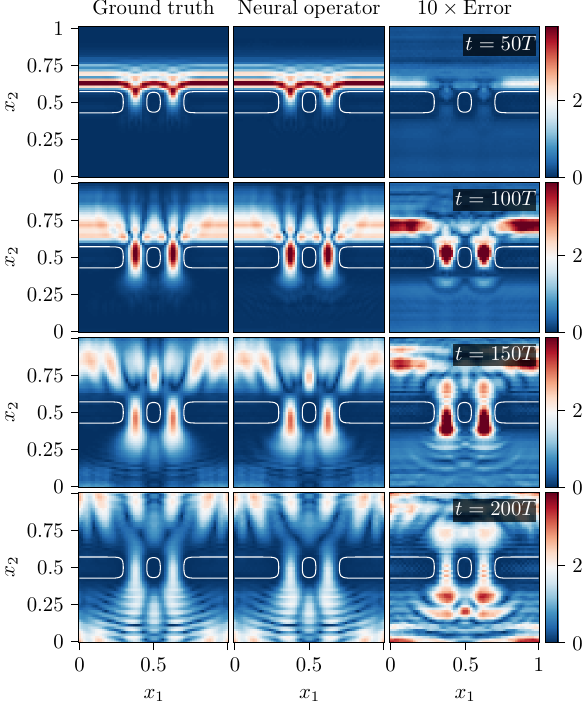}
	\caption{\label{fig:double-slit} Double-slit experiment in $d=2$. \textbf{From left to right:} Absolute values $|\Psi_{\GT}(\vec{x},t)|$; $|\Psi_{\NO}(\vec{x},t)|$; and $10 \times |\Psi_{\GT}(\vec{x},t) - \Psi_{\NO}(\vec{x},t)|$ for the indicated times $t$. An outline of the potential is plotted with white lines.}
\end{figure}

\paragraph{\normalfont\textbf{Performance}}
As before, we construct $\N^k$ to study time evolution of the wave function. The results are illustrated in Fig.~\ref{fig:double-slit}.
After passing through the double-slit openings, the wave packet interferes with itself, creating a fringe pattern in the $x_1$-direction. Due to periodic boundary conditions, it also develops interference in the $x_2$-direction at late times. As in the $d=1$ case, the numerical errors build up over time: in the first and last time-frames, $t = T$ and $t = 256T$, the $L^2$-error between the exact and inferred computations is $\sim 5 {\times} 10^{-4}$ and $\sim 0.14$ respectively.

\paragraph{\normalfont\textbf{Zero-shot super-resolution}} 
Let us finish this discussion by emphasizing the distinction between a neural network acting on a very large but finite-dimensional Hilbert space (obtained by a specific discretization of $D$) and a neural operator, which acts on an infinite-dimensional space (independent of the discretization). In particular, it means that the latter can be trained on a lower-resolution lattice and then applied to computing $\Psi_{\NO}(\vec{x}, T)$ at higher resolution. In the literature, this procedure is called a \emph{zero-shot super-resolution} \cite{kovachki2021neural}.

To exemplify it, in Tab.~\ref{tab:performance} we collected indicative times for computing $\Psi_{\GT}(\vec{x},T)$ and $\Psi_{\NO}(\vec{x},T)$ from the testing set at various resolutions, together with the $L^2$-errors between the two techniques. Finite-difference methods scale badly at large lattice sizes because they involve computing and taking powers of an $N^d \times N^d$ matrix for every new potential $V$. On the other hand, neural operator trained on the original sizes (here, $N=256$ in $d=1$ and $N=64$ in $d=2$) can be applied to finer grids and new potentials with low overhead.

\section{Discussion}

In this work, we explored the idea of using deep operator learning as a computational tool for mapping between function spaces appearing in fundamental physics. As an illustrative example, we considered neural operators predicting quantum-mechanical scattering of wave packets with potential barriers and wells. We envisage that neural operators have much broader applicability in quantum theory, beyond just time-evolution operators, both in numerical and symbolic manipulations.

Let us emphasize that this strategy depends on a reliable way of producing training samples. The prospect is that, once trained, a neural operator can perform the same computation approximately but much more efficiently. The biggest open question concerns out-of-distribution errors. For example, given that neural operators represent physics in unconventional ways, would cheaply-obtainable training data (e.g., coming from exactly-solvable systems) suffice to learn solutions to conventionally difficult problems? We leave a study of this provocative question until future work.

\begin{table}
	\begin{tabular}{c|c||c|c||c|c||c}
		$d$ & $N$ & CN time & CN mem. & NO time & NO mem. & Error \\
		\hline
		\underline{$1$} & \underline{$256$} & 0.013 & 0.004 & 0.002 & 0.13 & 0.0005 \\
		$1$ & $2048$ & 0.2 & 0.3 & 0.003 & 0.14 & 0.004 \\
		$1$ & $16384$ & 56 & $20$ &  0.003 & 0.2 & 0.004 \\
		$1$ & $32768$ & $\times$ & $>40$ &  0.003 & 0.3 & $\times$ \\
		\underline{$2$} & \underline{$64$} & 0.5 & $0.6$ & 0.005 & 0.16 & 0.0011 \\
		$2$ & $128$ & 22 & $10$ & 0.005 & 0.2 & 0.03 \\
		$2$ & $256$ & $\times$ & $>40$ & 0.005 & 0.4 & $\times$ 
	\end{tabular}
	\caption{\label{tab:performance}Comparison between single time-step GPU times (in seconds) and memory usage (in GB) of the Crank--Nicholson (CN) method and neural operator (NO) for different dimensions $d$ and lattice sizes $N$. The last column gives an average $L^2$-error between the two methods across the testing dataset. Original training sizes for NO are underlined. Crosses indicate that a computation did not terminate due to memory shortage.}
\end{table}

\section{Acknowledgments}

The author thanks Nima Arkani-Hamed, Carolina Figueiredo, Aidan Herderschee, Aaron Hillman, and the participants of the S-Matrix Bootstrap Workshop at the SwissMAP Research Station in Les Diablerets for useful comments.
The author gratefully acknowledges funding provided by the Sivian Fund and the grant DE-SC0009988 from the U.S. Department of Energy.

\bibliography{references}
	
\end{document}